\documentclass{ptephy}%%%%where ptephy is the template name

\usepackage{ulem} %% for strike-through
\usepackage[usenames]{color}

\newcommand{\ltk}{\left\{ }
\newcommand{\rtk}{ \right\} }
\newcommand{\ldk}{\left[ }
\newcommand{\rdk}{ \right] }
\begin{document}

\title{Complex 2D Matrix Model and Geometrical Map on Complex-$N_c$ Plane}

%\author{\name{First Author}{1}, \name{Second Author}{2,\dag}, and \name{Third Author}{3,\ast,}\thanks{These authors contributed equally to this work.}}
\author{\name{Kanabu Nawa}{1, \ast}, \name{Sho Ozaki}{1,2}, \name{Hideko Nagahiro}{3,4}, \name{Daisuke Jido}{5,6}, and \name{Atsushi Hosaka}{4}}
%%%%%%%%%%% The \name command should be used as \name{Insert author name here}{Insert affiliation number here}
%%%%% Please use \thanks for contributed author details

%%%%%%%%%%% The \affil command should be used as \affil{Insert affiliation number here}{Insert author address here}
\address{\affil{1}{Quantum Hadron Physics Laboratory, RIKEN Nishina Center, Saitama 351-0198, Japan}
\affil{2}{Institute of Physics and Applied Physics, Yonsei University, Seoul 120-749, Korea}
\affil{3}{Department of Physics, Nara Women's University, Nara 630-8506, Japan}
\affil{4}{Research Center for Nuclear Physics (RCNP), Osaka University, Osaka 567-0047, Japan}
\affil{5}{Yukawa Institute for Theoretical Physics, Kyoto University, Kyoto 606-8502, Japan}
\affil{6}{J-PARC Branch, KEK Theory Center, Institute of Particle and Nuclear Studies,
%High Energy Accelerator Research Organization (KEK),
%203-1, Shirakata, Tokai, 
Ibaraki 319-1106, Japan}
\email{knawa@riken.jp}}

\begin{abstract}%
We study the parameter dependence of the internal structure of
resonance states
by formulating Complex two-dimensional (2D) Matrix Model,
where the two dimensions represent two-levels of resonances.
We calculate a critical value of the parameter at which  
``nature transition''  with character exchange occurs between two resonance states,
from the viewpoint of geometry on complex-parameter space.
Such critical value is useful to know the internal structure of resonance states
with variation of the parameter in the system.
We apply the model to 
analyze the internal structure of hadrons
with variation of the color number $N_c$ from $\infty$ to a realistic value $3$.
By regarding $1/N_c$ as the variable parameter in our model,
we calculate a critical color number of nature transition between hadronic states
in terms of quark-antiquark pair and mesonic molecule as exotics
from the geometry on complex-$N_c$ plane.
For the large-$N_c$ effective theory,
we employ the chiral Lagrangian induced by holographic QCD with D4/D8/$\overline{\rm D8}$
multi-D brane system in the type IIA superstring theory.
\end{abstract}

\subjectindex{D32}

\maketitle

\section{Introduction}
\begin{figure}[t]
  \begin{center}
%   \hspace*{30mm}
       \resizebox{97mm}{!}{\includegraphics{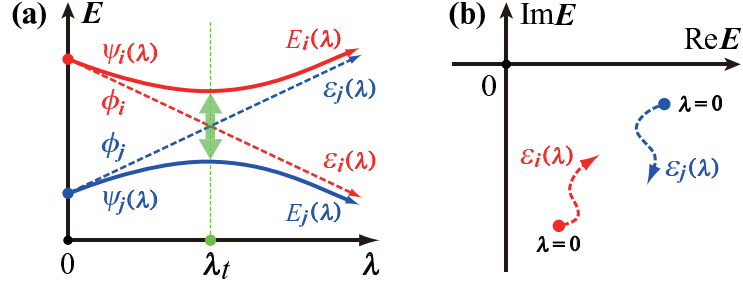}}\\
%       \resizebox{50mm}{!}{\includegraphics{fig_7_30.eps}}\\
  \end{center}
%\vspace{-4mm}
\caption{(Color) (a) Anticrossing between $i$th and $j$th eigenstates 
of Hermite Hamiltonian $\hat{H}(\lambda)$ with variation of $\lambda\in\mbox{${\bf R}$}$.
Indices of lines are explained in the text.
$i$th and $j$th eigenstates exchange their characters at anticrossing point $\lambda=\lambda_t$
as the nature transition.
(b) $\varepsilon_i(\lambda)$ and $\varepsilon_j(\lambda)$
of non-Hermite Hamiltonian $\hat{H}(\lambda)$ 
with variation of 
$\lambda\in\mbox{${\bf R}$}$ on complex energy plane.
$\varepsilon_i(\lambda)$ and $\varepsilon_j(\lambda)$ generally have no degeneracy at certain value of $\lambda\in\mbox{${\bf R}$}$
except for accidental case.}
  \label{fig_1D_2D}
\end{figure}

{\it How  do characters of states
change with variation of a parameter
which specifies the property of the system or of the environment 
where the system is placed?}
This is a general issue discussed in various phenomena of physics;
deformed nuclei depending on 
the deformation parameter of nuclear mean-field potential~\cite{RingS}, 
electronic wave function configurations of diatomic molecules 
depending on the internuclear distance~\cite{Molecule},
and conversion of solar neutrinos
depending on the distance from the center of the sun~\cite{Be}.
In quantum mechanics, one  starts with a Hermite model Hamiltonian
$\hat{H}(\lambda)$ with a real parameter $\lambda$.
Here one can assume that
the eigenstates 
$\phi_i$ ($i=1,2,\cdots$)
of $\hat{H}(\lambda)$ at $\lambda=0$
can be an appropriate basis with clear characters
to classify the properties of the eigenstates 
$\psi_i(\lambda)$ ($i=1,2,\cdots$)
for finite $\lambda$.
%in terms of the characters of $\phi_i$.
%
Now, if the energy expectation values 
$\varepsilon_i(\lambda)\equiv \langle \phi_i |\hat{H}(\lambda)|\phi_i \rangle$
($i=1,2,\cdots$) cross with each other at a certain value
$\lambda=\lambda_t\in \mbox{${\bf R}$}$,
the energy eigenvalues $E_i(\lambda)$ of $\psi_i(\lambda)$ with finite mixing
have level repulsion, i.e., anticrossing at $\lambda_t$ (see Fig.~\ref{fig_1D_2D}(a))
due to the Neumann-Wigner non-crossing rule~\cite{NW}.
At this point,
the overlap between $\psi_i(\lambda)$ and $\phi_i$ is exceeded by that
between $\psi_i(\lambda)$ and $\phi_j$ ($i\neq  j$)
as $|\langle\phi_i|\psi_i\rangle|^2\leq |\langle\phi_j|\psi_i\rangle|^2$.
Therefore, due to orthogonality,  $\psi_i(\lambda)$ and $\psi_j(\lambda)$ 
exchange their characters
in terms of the appropriate basis $\phi_i$ and $\phi_j$
at the anticrossing point $\lambda=\lambda_t$,
which we call ``nature transition'' in this paper. 
In fact, 
the critical value $\lambda_t$
is very useful to know 
the internal structure of the quantum states
with variation of certain parameter $\lambda$.

In this paper, we consider 
the quantum systems with dissipation into 
decay channels outside of the model space.
Such systems are often called {\it open quantum systems} with resonance states,
which are effectively described by a non-Hermite model Hamiltonian 
$\hat{H}(\lambda)$ with complex energy eigenvalues~\cite{Moiseyev}.
The real and imaginary parts of the eigenvalues correspond to the mass and decay width
of the resonance states, respectively.
%
%\corr{}{Their non-Hermitian nature is essentially same as that of the 
%effective Hamiltonian in the 
%Feshbach reduction formalism~\cite{Feshbach, Rotter, Hatano_F}.}
It can be shown in the Feshbach reduction formalism that the Hamiltonian
with a reduced model space becomes non-Hermitian~\cite{Feshbach, Rotter, Hatano_F}. 
In such open quantum systems,
$\varepsilon_i(\lambda)$ of $\phi_i$
can move on the complex energy plane (see Fig.~\ref{fig_1D_2D}(b))
without having degeneracy at a certain value of $\lambda$ except for an accidental case~\cite{degene}.
Therefore, a simple criterion 
should be newly found to 
judge the existence of the nature transition between the resonances, and 
%calculate 
its critical value $\lambda_t$
can be used to know the internal structure of the resonance states depending 
on the parameter.
In this paper, we construct
Complex two-dimensional (2D) Matrix Model 
to discuss the nature transition between
two resonance states.
(Two dimensions represent two levels of resonances.)
This 2D model will
give an elementary understanding 
for higher dimensional problems 
because the latter can often be reduced locally to the 2D problems.
We show that, by extending $\lambda$ to a {\it complex variable},
the geometry
on the complex-$\lambda$ plane gives a criterion of the nature transition
within the real parameter subspace 
$\lambda\in \mbox{${\bf R}$}$.

%---------------------------------------------------------------------
%-----------------------------------------------------------------------

After establishing the general framework,
we apply it to 
the hadron physics with strong interaction, which is governed by
{\it quantum chromodynamics} (QCD)
as the ${\rm SU}(N_c)$ gauge theory 
with color number $N_c=3$~\cite{Nambu}.
By extending $N_c$ to an arbitrary number,
$1/N_c$-expansion 
provides
a systematic perturbative treatment.
The leading order of ``large-$N_c$ QCD'' 
reproduces lots of QCD phenomenologies~\cite{tH, Witten}.
In fact, 
in large-$N_c$ QCD,
the internal structure of mesons becomes clear:
mesons as quark-antiquark ($q\bar{q}$) pairs
appear with masses of $O(N_c^0)$ and zero widths,
while
``mesonic molecules'' 
can also appear as resonances
with masses and widths increasing along with $N_c$ because the meson-meson
interactions are suppressed with $O(1/N_c)$~\cite{PR}.
From such
considerations in large-$N_c$,
one often expects that exotics can also be suppressed in the real world~\cite{Witten}.
However 
the internal structures of hadrons can be easily changed due to the development of hadron dynamics
scaled by $1/N_c$.
Here, a basic but essential question arises: 
{\it what is the internal structure of hadrons 
with continuous variation of $N_c$ from $\infty$ to $3$?}
To find a typical feature for such $N_c$-dependence of the internal structure of hadrons,
we adopt the Complex 2D Matrix Model.
By regarding $q\bar{q}$ and mesonic molecule states in large-$N_c$ 
as the appropriate basis 
$\phi_i$ ($i=1,2$) with clear characters, 
and by identifying 
$1/N_c$ to $\lambda$ in the Complex 2D Matrix Model,
we will calculate a critical color number of nature transitions 
in terms of appropriate basis from the geometry on the complex-$N_c$ plane.
As an example,
we investigate the internal structure of $a_1(1260)$ meson
with admixed nature of $q\bar{q}$ and $\pi\rho$-molecule components.
For the large-$N_c$ effective theory,
we employ the chiral Lagrangian induced by holographic QCD with D4/D8/$\overline{\rm D8}$
multi-D brane system in the type IIA superstring theory~\cite{SS, NSK}.

In Sec.~\ref{C2D},
we formulate Complex 2D Matrix Model.
In Sec.~\ref{CNc},
we discuss the application of the model to
$N_c$-dependence of internal structure of hadrons.
Sec.~\ref{Sum}
is devoted to summary and outlook.
In Appendix~\ref{A1},
we calculate the attaching number $N_{\rm at}$
which characterizes  the geometry near the origin on the
complex-parameter space.
In Appendix~\ref{A2},
we show a simple prescription of writing geometrical maps 
for arbitrary matrix elements of the model.

%%%%%%%%%%%%%%%%%%%%%%%%%%%%%%%%%%%%%%%%%%%%%%%%%%%%%%%
\section{Complex 2D Matrix Model}\label{C2D}
%%%%%%%%%%%%%%%%%%%%%%%%%%%%%%%%%%%%%%%%%%%%%%%%%%%%%%%
First we formulate
Complex 2D Matrix Model 
to treat a two-level problem
in a quantum system with resonances.
We describe resonance states 
by using the bi-orthogonal representation
as $|\phi_i)$$(i=1,2)$: 
its bra-state is defined by the complex conjugate 
of the Dirac bra-state $(\phi_i|\equiv\langle \phi_i^*|$,
which was firstly introduced for the unstable nuclei in nuclear physics~\cite{Ho, Berggren, Romo}.
Only by taking such bi-orthogonal representation,
resonance states with different eigenvalues become orthogonal to each other as
$(\phi_i|\phi_j)=\delta_{ij}$, 
which is needed to employ the matrix representation of operators in such basis. 
As anticipated, we assume that $|\phi_i)$, the eigenstates of $\hat{H}(\lambda)$ 
at $\lambda=0$,
are the appropriate basis with clear characters and are
useful to classify the quantum states.
Hence we consider the Hamilton matrix 
${H}(\lambda)\equiv [(\phi_i|\hat{H}(\lambda)|\phi_j)]$ 
in this basis:
\begin{eqnarray}
H(\lambda)=
                  \left(\begin{array}{cc}
                   \varepsilon_1(\lambda) & V_{12}(\lambda) \\
                    V_{21}(\lambda)               & \varepsilon_2 (\lambda)
                             \end{array}\right),
\label{H_1}
\end{eqnarray}
where $\varepsilon_i(\in\mbox{${\bf C}$})$ is the energy of $|\phi_i)$ 
and $V_{ij}(\in \mbox{${\bf C}$})$ are the interaction
satisfying
%$V_{12}=V_{21}$ and 
$V_{ij}(0)=0$.
$\lambda (\in\mbox{${\bf R}$})$ is a parameter,
controlling the development of the two eigenstates 
$|\psi_i(\lambda))$
which can be obtained in terms of the basis  $|\phi_i)$ as
\begin{eqnarray}
|\psi_i(\lambda)) \equiv C_{i 1}(\lambda)|\phi_1)
                                                         + C_{i 2}(\lambda)|\phi_2). \hspace{3.5mm}(i=1,2) 
\label{DE1}
\end{eqnarray} 
The coefficients $C_{ij}(\lambda)$ 
carry the information for
the internal structure of the eigenstates $|\psi_i(\lambda))$ 
in terms of $|\phi_i)$.
There is a subtlety 
for the interpretation of component weights
from $C_{ij}(\lambda)$,  
since the norms $(\psi_i|\psi_i)=C_{i1}^2+C_{i2}^2$ 
can be complex numbers due to the bi-orthogonality.
Several attempts have been made to interpret
such complex probability of resonances 
(for example, see Ref.~\cite{Bergg-rep}),
while a consensus has not been achieved yet.
%\corr{}{
%Recently, Ref.~\cite{Hatano-rep} considers a probabilistic
%interpretation for a square modulus of resonance wave function
%in an expanding region; its boundary runs with a speed of leaking particles.
%However, as for the expansion coefficients in Eq.~(\ref{DE1}),
%their probabilistic interpretation still remains unsolved.
%}
%
Recently Ref.~\cite{Hatano-rep} has considered 
a probabilistic interpretation of resonance states 
 by taking the integral of the modulus square of
 resonance wave function over a limited spatial domain expanding with the speed
 of leaking particles. 
 Normalization of the resonance wave function
 over such domain makes the modulus square finite and could be
 suitable for the probabilistic interpretation.
However, as for the expansion coefficients in Eq.~(\ref{DE1}),
their probabilistic interpretation still remains unsolved.
In this work we simply presume the module, $|C_{ij}(\lambda)|^2$,  
to be interpreted as the component weights, as it is suitable for narrow resonances. 
At $\lambda=0$, $|\psi_i(\lambda))$ coincides with $|\phi_i)$
due to $V_{ij}(0)=0$, so that $C_{ij}(0)=0$ for $i\neq j$.

Now, if $\hat{H}(\lambda)$ is Hermite with real eigenvalues,
the level crossing of $|\phi_i)$
is known to give 
the level anticrossing of $|\psi_i(\lambda))$ as shown in Fig.~\ref{fig_1D_2D}(a)~\cite{NW}.
At this
anticrossing point $\lambda=\lambda_t$,
$|\psi_i(\lambda))$
exchange their characters
as ``nature transition''
with the transition condition 
$|C_{i 1}(\lambda_t)|^2=|C_{i 2}(\lambda_t)|^2$,
where the two basis components $|\phi_1)$ and $|\phi_2)$
are equally mixed 
%within the eigenstates of Eq.~(\ref{DE1})
as a character exchanging point.
In this paper, we newly consider 
the case that $\hat{H}(\lambda)$ is non-Hermite with complex eigenvalues for
resonance states.
As we will show below, $|C_{i 1}|^2=|C_{i 2}|^2$ can be 
satisfied {\it at least} by the energy coincidence $E_1=E_2$,
which can be realized if one extends $\lambda$ to a {\it complex variable}~\cite{Kato}.
Therefore,
to get a geometrical insight for the existence of 
nature transition, here
%To judge the existence of adiabatic transition between two-resonance states,
we introduce the {\it complex-$\lambda$ plane}. 

\begin{figure}[t]
  \begin{center}
       \resizebox{103mm}{!}{\includegraphics{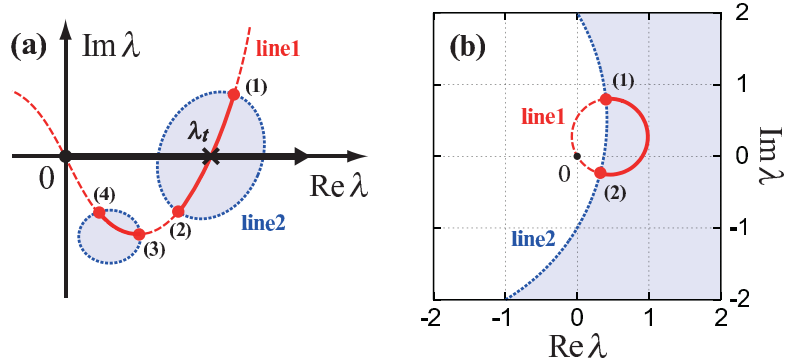}}\\
  \end{center}
%\vspace{-4mm}
\caption{(Color) (a) Schematic figure of geometrical map with
transition lines and 
exceptional points 
on complex $\lambda$ plane.
Line 1 and shaded area with boundary of line 2
correspond to the conditions (\ref{condF1}) and (\ref{condF2}),
respectively.
Points (n) denote the exceptional points $\lambda_{\rm EX}^{(n)}$.
Transition lines are shown by the solid curves, which satisfy both  (\ref{condF1}) and (\ref{condF2}).
%i.e., $|R_a(\lambda)|=1$.
(b) Linear-$\lambda$ model with 
$\varepsilon_1^{(0)}=1000-200i$,
$\varepsilon_2^{(0)}=1200$,
$v_{11}=100+100i$,
$v_{22}=-100-200i$,
$v_{12}=v_{21}=200+50i$ in MeV unit as test values.
%(c) Non-linear-$\lambda$ model with same parameter sets in (b).
}
  \label{fig_1}
\end{figure}

By solving the Schr\"{o}dinger equation: $\hat{H}|\psi) =E|\psi)$,
we find the two eigenvalues $E_i(\lambda)$  ($i=1, 2$) as
\begin{eqnarray}
E_i(\lambda) &=& 
        \{\varepsilon_1(\lambda)+\varepsilon_2(\lambda)\}/2
                \pm F(\lambda), \label{Eab_1}\\
F(\lambda)&\equiv& \sqrt{A(\lambda)^2+\overline{V}(\lambda)^2}, \label{Eab_2}\\
A(\lambda)&\equiv& \{\varepsilon_1(\lambda)-\varepsilon_2(\lambda)\}/2,\\
\overline{V}(\lambda)^2&\equiv&V_{12}(\lambda)V_{21}(\lambda),
                             %\equiv \varepsilon + V(\lambda). 
                             \label{Eab_3}
\end{eqnarray} 
and the coefficient ratios $R_i(\lambda)$ of the eigenstates $|\psi_i (\lambda))$ ($i=1, 2$)
in Eq.~(\ref{DE1}) as 
\begin{eqnarray}
R_i(\lambda)\equiv \frac{C_{i 2}(\lambda)}{C_{i 1}(\lambda)}
=-\frac{1}{V_{12}(\lambda)}\{A(\lambda)\mp F(\lambda)\}.\label{Rab_1}
\end{eqnarray}
The upper (lower) sign in Eqs.~(\ref{Eab_1}) and (\ref{Rab_1})
corresponds to $i=1$ ($i=2$).
The ratios (\ref{Rab_1})
are sufficient to discuss the nature transition between two levels as below.
Now we consider the transition condition 
$|C_{i 1}(\lambda)|^2=|C_{i 2}(\lambda)|^2$
on the complex-$\lambda$ plane.
Due to the bi-orthogonality $(\psi_1|\psi_2)=0$, i.e.,
$R_1 R_2=-1$,
the transition condition can be written only by the ratios (\ref{Rab_1}) as
%$R_a$ and $R_b$ as
%
$|R_1(\lambda)|=|R_2(\lambda)|$,
which is equivalent from Eq.~(\ref{Rab_1}) to
\begin{eqnarray}
{\rm Re}[A(\lambda)^*F(\lambda)]=0.\label{cond3}
\end{eqnarray}
Due to the square root in (\ref{Eab_2}), Eq.~(\ref{cond3}) becomes equivalent to the two conditions:
\begin{eqnarray}
&&{\rm Re}[A(\lambda)^*\overline{V}(\lambda)]=0,\label{condF1}\\
&&|A(\lambda)|^4-\{{\rm Im}[A(\lambda)^*\overline{V}(\lambda)]\}^2 \leq 0. \label{cond4}
\end{eqnarray}
From Eq.~(\ref{condF1}), 
$|{\rm Im}[A(\lambda)^* \overline{V}(\lambda)]|=|A(\lambda)||\overline{V}(\lambda)|$,
so that the condition (\ref{cond4}) becomes  
\begin{eqnarray}
|A(\lambda)|^2\leq|\overline{V}(\lambda)|^2, \label{condF2}
\end{eqnarray}
which has been divided by $|A(\lambda)|^2$
since $\lambda$'s for $A(\lambda)=0$ trivially 
satisfy the conditions (\ref{condF1}) and (\ref{condF2}).
Now the ``transition line'' is defined 
as the region satisfying 
$|C_{i 1}(\lambda)|^2=|C_{i 2}(\lambda)|^2$, i.e.,
both conditions (\ref{condF1}) and (\ref{condF2})
on the complex-$\lambda$ plane.
Therefore, the line (\ref{condF1}), named ``line 1'', 
can be the candidate of the transition line,
and the region (\ref{condF2}) with the boundary 
$|A(\lambda)|^2=|\overline{V}(\lambda)|^2$, named ``line 2'',
selects the proper part for the transition line.
%
%Here it is noted that
The region (\ref{condF2}) always excludes the origin $\lambda=0$
for the case $\varepsilon_1(0)\neq \varepsilon_2(0)$,
because $|A(0)|> 0$ and $|\overline{V}(0)|=0$. 
%$|V_{12}(0)|=0$.
%
Then, if the transition line crosses the real-$\lambda$ axis,
the nature transition occurs at the crossing point  
$\lambda=\lambda_t\in \mbox{\bf R}$
(see schematic Fig.~\ref{fig_1}(a)).

Now, from (\ref{condF1}) and (\ref{condF2}),
the crossing points $\lambda=\lambda_{\rm EX}^{(n)}\in \mbox{\bf C}$ ($n=1,2,\cdots$)
of line 1 and line 2
%and the boundary of the region (\ref{condF2})
satisfy the condition
$A(\lambda)^*F(\lambda)=0$
for $^{\forall}A(\lambda)\neq 0$,
which is equivalent to
\begin{eqnarray}
F(\lambda)^2=A(\lambda)^2+\overline{V}(\lambda)^2=0. \label{EXcond1}
\end{eqnarray}
Therefore, at $\lambda=\lambda_{\rm EX}^{(n)}$, 
the mass difference in Eq.~(\ref{Eab_1}) becomes zero and 
two eigenvalues
coincide 
%with each other 
as $E_1(\lambda)=E_2(\lambda)$.
$\lambda_{\rm EX}^{(n)}$ are
called the ``exceptional points'' on the complex-$\lambda$ plane~\cite{Kato}.
In fact, the importance of the exceptional points has been
intensively studied both theoretically~\cite{QC} and experimentally~\cite{QCexperiment} 
in the area of quantum chaos,
where the dense exceptional points on the complex-$\lambda$ plane
correspond to the development of quantum chaos in the energy-level statistics~\cite{QC}.
Now, in this paper, we can show that
line 1 and line 2 cross each other at all exceptional points,
so that these points
can always be the {\it end points}
of the transition lines.
Therefore,
the location of the exceptional points 
is very important to geometrically judge the existence of $\lambda_t\in \mbox{\bf R}$. 
%

%---------------------------------------------------------------------
%-----------------------------------------------------------------------

\begin{figure}[t]
  \begin{center}
       \resizebox{88mm}{!}{\includegraphics{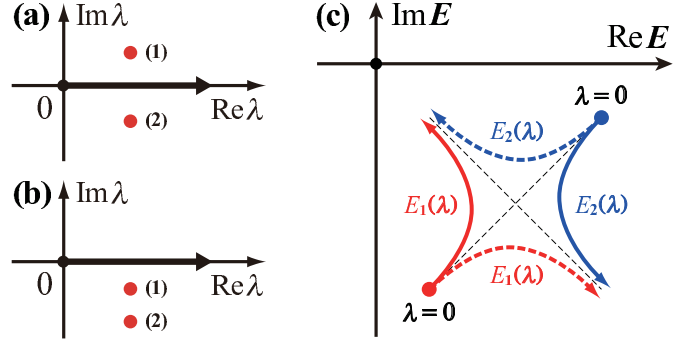}}\\
  \end{center}
%\vspace{-4mm}
\caption{(Color) Two exceptional points on complex-$\lambda$ plane;
(a) two locate on the opposite side striding over the real-$\lambda$ axis, 
and (b) two locate without striding it.
Bold arrow implies the variation
of $\lambda$ within real parameter subspace. 
(c) Eigenvalue behaviors on the complex energy plane with 
level anticrossing/width crossing shown by the solid lines and
level crossing/width anticrossing shown by the dashed lines.
}
  \label{fig_two_excep}
\end{figure}
%-----------------------------------------------------------------------

One simple example is the  
``linear-$\lambda$ model'' with Hamilton matrix:
\begin{eqnarray}
H(\lambda)=
                  \left(\begin{array}{cc}
                   \varepsilon_1^{(0)} & 0\\
                     0        & \varepsilon_2^{(0)}
                             \end{array}\right)+
\left(\begin{array}{cc}
                   \lambda v_{11}   & \lambda v_{12}\\
                   \lambda v_{21}   & \lambda v_{22}
                             \end{array}\right),
\label{Linear-l}
\end{eqnarray}
where $\varepsilon_i^{(0)}$ and $v_{ij}$ ($i,j=1,2$) are $\lambda$-independent quantities.
Two exceptional points and one transition line appear
(see Fig.~\ref{fig_1}(b)), which are simply checked from the power counting about $\lambda$ 
in (\ref{EXcond1}).
In this model, Eq.~(\ref{condF1}) can be equally written 
with $\Delta\varepsilon\equiv \varepsilon_1^{(0)}-\varepsilon_2^{(0)}$, 
$\Delta v \equiv v_{11} -v_{22}$, and
$\overline{v}^2\equiv v_{12}v_{21}$ as
\begin{eqnarray}
|\lambda-\overline{\lambda}|=|\overline{\lambda}|, 
\hspace{2mm}(\overline{\lambda}\equiv -(\Delta\varepsilon) \overline{v}^* \{2 {\rm Re}[(\Delta v)\overline{v}^* ]\}^{-1}) \label{circle1}
\end{eqnarray} 
so that line 1 
is a circle crossing at 
$\lambda=0$, $\lambda_{\rm EX}^{(1)}$ and $\lambda_{\rm EX}^{(2)}$,
and
the transition line has an arc shape.
Ref.~\cite{QC} shows that, in the linear-$\lambda$ model,
the eigenvalue behaviors for $\lambda \in \mbox{\bf R}$
depend on the location of the two exceptional points;
if the two locate on the opposite sides striding over the real-$\lambda$ axis (see in Fig.~\ref{fig_two_excep}(a)),
level anticrossing/width crossing occurs, 
while, if not (see Fig.~\ref{fig_two_excep}(b)),
level crossing/width anticrossing occurs as in Fig.~\ref{fig_two_excep}(c).
Therefore, 
by comparing Fig.~\ref{fig_1}(b) 
and Fig.~\ref{fig_two_excep},
we can newly  suggest that
the nature transition occurs only in the level anticrossing/width crossing case.
In this way, 
as for the linear-$\lambda$ model with two exceptional points,
we can relate the behaviors of poles on the complex-energy plane
and their internal structures through the geometry on the complex-$\lambda$ plane.
The linear-$\lambda$ model also suggests that,
if $v_{11}=v_{22}=0$, the radius of the circle of line 1 
in Eq.~(\ref{circle1}) diverges: $|\overline{\lambda}|\rightarrow \infty$,
so that there is no nature transition for finite $\lambda$.
There only occurs the mixing of the basis components up to $50\%$ at most.
Therefore the $\lambda$-dependence in the diagonal components
of the matrix form (\ref{H_1}) is needed to have the 
nature transition between resonance states.
%

%-----------------------------------------------------------------------

\begin{figure}[t]
  \begin{center}
       \resizebox{154mm}{!}{\includegraphics{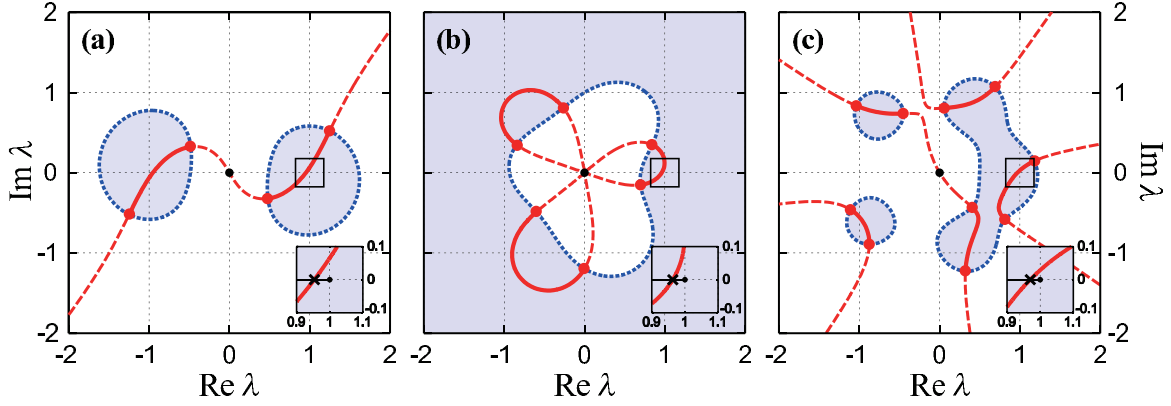}}\\
  \end{center}
%\vspace{-4mm}
\caption{(Color) Geometrical maps of 
(a) $H_{221}$, 
(b) $H_{233}$ and 
(c) $H_{151}$  with 
$\varepsilon_1^{(0)}=1000-200i$, 
$\varepsilon_2^{(0)}=1200$,
$v_{11}=100+100i$,
$v_{22}=-100-200i$,
$v_{12}=v_{21}=200+50i$
in MeV unit as test values.
Notations for line 1, line 2, exceptional points and transition lines
are same as those in Fig.~\ref{fig_1}(a). 
Blank squares show the area around $\lambda=1$.
}
  \label{PQR_fig}
\end{figure}
%-----------------------------------------------------------------------

We can show more general cases of the geometrical map as the
``$PQR$ model'' by using the following matrix:
\begin{eqnarray}
H(\lambda)_{PQR}=
                  \left(\begin{array}{cc}
                   \varepsilon_1^{(0)} & 0\\
                     0        & \varepsilon_2^{(0)}
                             \end{array}\right)+
\left(\begin{array}{cc}
                   \lambda^P v_{11}   & \lambda^R v_{12}\\
                   \lambda^R v_{21}   & \lambda^Q v_{22}
                             \end{array}\right),
\label{PQRmodel}
\end{eqnarray}
where $P,Q$ and $R$ are the powers of $\lambda$ in the matrix elements.
As examples, geometrical maps of $H_{221}, H_{233}$ and $H_{151}$
with $\varepsilon_i^{(0)}$ and $v_{ij}$ fixed
are presented in Fig.~\ref{PQR_fig}.
By changing the values of $P,Q$ and $R$,
various types of geometry can be observed on the complex-$\lambda$ plane.
Fig.~\ref{PQR_fig} and Fig.~\ref{fig_1}(b) classified as
$H_{111}$ also show 
that the geometry around $\lambda=1$ is rather 
independent of the values of $P, Q$ and $R$ 
with the same values of 
$\varepsilon_i^{(0)}$ and $v_{ij}$.
%\corr{which is simply because 
%any power of unity is unity.}{}
%\corr{Such power-independence around $\lambda=1$ is one of the interesting features
%of the geometrical maps.}{
On the other hand, $H_{001}$ effectively
corresponds to the linear-$\lambda$ model with $v_{11}=v_{22}=0$,
so that there is no nature transition for finite $\lambda$ as discussed
below Eq.~(\ref{circle1}).
(If $\varepsilon_i^{(0)}$ and
$v_{ij}$ ($i,j =1,2$) are specially chosen for line 1 to exactly
cross the point $\lambda=1$,
$H_{001}$ still corresponds to the  linear-$\lambda$ model with
$v_{11}=v_{22}=0$
where line 1 as a circle of infinite radius coincides with the real-$\lambda$
axis itself.)
These features can be roughly understood as follows;
First, let us consider a situation that
line 1 of Eq.~(\ref{condF1})
appears near $\lambda=1$,
i.e., Eq.~(\ref{condF1}) is satisfied at
$\lambda=1+\xi$ (${}^\exists\xi\in \mbox{\bf C}, |\xi|\ll 1$) as
\begin{eqnarray}
{\rm Re}\ldk\ltk\Delta\varepsilon + (1+\xi)^P v_{11}-(1+\xi)^Q
 v_{22}\rtk^\ast(1+\xi)^R\bar{v}\rdk=0,\label{formxi1}
\end{eqnarray}
with $\Delta\varepsilon\equiv\varepsilon_1^{(0)}-\varepsilon_2^{(0)}$
and $\bar{v}^2\equiv v_{12}v_{21}$. 
LHS of Eq.~(\ref{formxi1})
can be expanded up to $O(\xi^1)$ as
\begin{eqnarray}
{\rm Re}\ldk (\Delta\varepsilon + \Delta v)^\ast \bar{v}  \rdk
+{\rm Re}\ldk \xi^\ast v_{11}^\ast \bar{v}  \rdk P
-{\rm Re}\ldk \xi^\ast v_{22}^\ast \bar{v}  \rdk Q
+{\rm Re}\ldk (\Delta\varepsilon + \Delta v)^\ast  \xi \bar{v}  \rdk R =0, \label{formxi2}
\end{eqnarray}
with $\Delta v=v_{11}-v_{22}$.
Eq.~(\ref{formxi2}) implies that,
with increase of $|P|$, $|Q|$ and $|R|$,
%as $|P|$, $|Q|$ and $|R|$ increase,
$|\xi|$ tends to decrease to maintain Eq.~(\ref{formxi2}).
That is, a part of line 1 approaches asymptotically to
$\lambda=1$ as its ``fixed point''.
Actually, $|P|$, $|Q|$ and $|R|$ determine, via Eq.~(\ref{formxi2}),
the order of $|\xi|$.
Furthermore,
line~1 always crosses the point $\lambda=0$ for $R>0$ 
(see Appendix~\ref{A1}),
so that $|\xi|$ should be {\it at least $O(1)$ or less} for any powers with $R>0$.
Therefore, $|P|$, $|Q|$ and $|R|$ 
which are {\it sufficiently larger than unity}
tend to develop a power-independent geometry of line 1
in the vicinity of $\lambda=1$ 
in comparison to a length scale $O(1)$ on the complex-$\lambda$ plane.

So far, we have formulated the model
described by the Hamiltonian 
 (\ref{H_1}) with
arbitrary complex functions:
$\varepsilon_i(\lambda)$ and $V_{ij}(\lambda)$,
%to treat the general 
and studied two-level problems on the complex-energy plane.
We have shown that
the geometrical map on the complex-$\lambda$ plane
provides the geometrical insight for the existence of nature transition
within the real parameter subspace $\lambda \in \mbox{\bf R}$.
For convenience,
we supply in Appendix~\ref{A2}
a prescription of writing geometrical maps 
for arbitrary $\varepsilon_i(\lambda)$ and $V_{ij}(\lambda)$, 
instead of numerically solving the high-powered 
algebraic equations (\ref{condF1}) and (\ref{condF2}).
%\newpage
%\vspace*{220mm}
%%%%%%%%%%%%%%%%%%%%%%%%%%%%%%%%%%%%%%%%%%%%%%%%%%%%%%%
\section{Application to $N_c$-dependence of internal structure of hadrons}\label{CNc}
%%%%%%%%%%%%%%%%%%%%%%%%%%%%%%%%%%%%%%%%%%%%%%%%%%%%%%%

Let us now  utilize the Complex 2D Matrix Model to 
find the typical $N_c$-dependence of the internal structure of hadrons.
As a demonstration, we consider
the $a_1(1260)$ meson which has 
admixed nature of $q\bar{q}$ and $\pi\rho$-molecule components.
First, we prepare the appropriate basis for the $q\bar{q}$ and the $\pi\rho$-molecule
states in large-$N_c$.
For the large-$N_c$ effective theory,
we make use of  the chiral Lagrangian 
induced by
holographic QCD
with D4/D8/$\overline{\mbox{D8}}$ multi-D brane system in the type IIA
superstring theory~\cite{SS, NSK}.
Due to the large-$N_c$ condition of the duality with
``classical'' supergravity,
the $a_1$ meson appearing as a gauge field in holographic QCD
should correspond to
the $q\bar{q}$ state.
On the other hand,
the holographic action also induces the energy-dependent
$\pi$-$\rho$ interaction as the Weinberg-Tomozawa (WT) interaction of
order $O(1/N_c)$. 
%as shown in Fig.~\ref{fig_diagram}(b).
%
Due to its attractive interaction,
the non-perturbative $\pi$-$\rho$ dynamics
gives a resonance pole as the ``$\pi\rho$-molecule state''.
%even for large-$N_c$, i.e., weak coupling region~\cite{Landau}.
%
The $a_1$ meson as the $\pi\rho$-molecule
is also studied in the chiral unitary model~\cite{M_1, M_2}. 
Thus, by preparing
the $q\bar{q}$ and $\pi\rho$-molecule states 
%at large-$N_c$
as the appropriate basis $\phi_i$ ($i=1,2$)
and identifying $1/N_c$ to $\lambda$ 
in the Complex 2D Matrix Model,
we will calculate the critical color number of the nature transition
from the geometry on the complex-$N_c$ plane.

\begin{figure}[t]
  \begin{center}
\hspace*{-5mm}
       \resizebox{57mm}{!}{\includegraphics{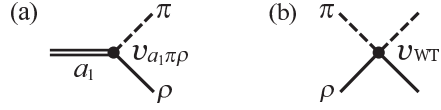}}\\
  \end{center}
%\vspace{-4mm}
\caption{Interactions between $\pi$, $\rho$ and $a_1$ mesons;
(a) three-point interaction and (b) Weinberg-Tomozawa interaction.
}
\label{fig_diagram}
\end{figure}

Here we
investigate the scattering equation for the $\pi$-$\rho$ propagator in  the 
$J^P=1^+$ channel.
By reducing the relativistic eigenvalue equation to the Schr\"{o}dinger
equation of the model (\ref{H_1}) with a non-relativistic approximation as discussed below, we will derive the geometrical map
on the complex-$N_c$ plane for the $a_1$ meson.
From the Lagrangian in holographic QCD~\cite{SS, NSK},
we obtain the three-point interaction
$v_{a_1\pi\rho}$ and the WT interaction $v_{\rm WT}$ 
in Fig.~\ref{fig_diagram} 
after proper $s$-wave projection~\cite{NNOJH} in the form,
\begin{eqnarray}
&&\hspace{-10mm}v_{a_1\pi\rho}\!=\!\frac{2\sqrt{2}}{f_\pi}g_{a_1\pi\rho}(s-m_\rho^2),  \label{3pc}    \\
&&\hspace{-10mm}v_{\rm WT}\!=\!- \frac{1}{4 f_\pi^{2}}\{3s\!-\!2(m_\rho^2+m_\pi^2)\!-\!(m_\rho^2-m_\pi^2)^2\frac{1}{s}\}.\label{v_WT}
\end{eqnarray}
By taking the two experimental inputs, e.g., $f_\pi=92.4$MeV and $m_\rho=776$MeV,
all the masses and coupling constants of hadrons can be uniquely determined
in the holographic approach as
$m_{a_1}=1189$MeV 
and 
$g_{a_1\pi\rho}=0.26$~\cite{SS}.
(In the D4/D8/$\overline{\mbox{D8}}$ model, 
pion is massless, 
whereas we use an isospin-averaged mass value: $m_\pi=138$MeV.)

%------------------------------------
%----------------------------

Now we introduce a two-dimensional G-function
with $\pi\rho$ and $q\bar{q}$ channels, having $J^P=1^+$ 
%quantum number as
as the $a_1$ meson:
\begin{eqnarray}
G^{-1}&=&G_0^{-1}-V\\
             &=&\left(\begin{array}{cc}
                                         G_{\pi\rho} & 0 \\
                                                 0  & G_{a_1}
                                          \end{array}\right)^{-1}-
                                         \left(\begin{array}{cc}
                                                 v_{\rm WT} & v_{a_1\pi\rho} \\
                                                  v_{a_1\pi\rho} & 0
                                          \end{array}\right), \label{Gform1}
\end{eqnarray}
where $G_{a_1}\equiv(s-m_{a_1}^2)^{-1}$
%is a bare-$a_1$ propagator and 
is a propagator for the $q\bar{q}$ state as the $a_1$ meson and
$G_{\pi\rho}$ is $\pi\rho$ loop function~\cite{M_1} as
\begin{eqnarray}
G_{\pi\rho}\equiv i \int\frac{d^4 q}{(2\pi)^4}\frac{1}{(P-q)^2-m_\pi^2+i\epsilon}
                                                                                    \frac{1}{q^2-m_\rho^2+i\epsilon},\label{Gpirho}
\end{eqnarray}
with $P$ a total incident momentum as $P^2=s$. 
We use a dimensional regularization
with the natural condition~\cite{HJH} 
to avoid the effect of CDD pole in Eq.~(\ref{Gpirho}).
In fact, the loop integral of Eq.~(\ref{Gpirho}) 
%algebraically derived from 
appears in the scattering equation of
the T-matrix 
with the separable approximation for the interactions~\cite{M_1}.
Then one can sum up the diagonal component of the potential in Eq.~(\ref{Gform1}) as
\begin{eqnarray}
G^{-1} =\left(\begin{array}{cc}
                                         G_{\rm WT} & 0 \\
                                                 0  & G_{a_1}
                                          \end{array}\right)^{-1}-
                                         \left(\begin{array}{cc}
                                                                                 0 & v_{a_1\pi\rho} \\
                                                  v_{a_1\pi\rho} & 0
                                          \end{array}\right), \label{Gform2}
\end{eqnarray}
with $G_{\rm WT}^{-1}\equiv G_{\pi\rho}^{-1}-v_{\rm WT}$.
We numerically find that $G_{\rm WT}$
has single resonance pole above the $\pi\rho$ threshold as
\begin{eqnarray}
G_{\rm WT}=\frac{G_{\pi\rho}}{1-v_{\rm WT}G_{\pi\rho}}\equiv
                              \frac{Z(s)}{s-s_p}. \label{G_molecule1}
\end{eqnarray}
This pole appears due to non-perturbative dynamics between $\pi$ and $\rho$ through the
4-point coupling $v_{\rm WT}$, 
%as in Fig.~\ref{fig_diagram} (b),
so that we interpret $(s-s_p)^{-1}$ in Eq.~(\ref{G_molecule1})
as the propagator of ``$\pi\rho$-molecule state''
with a wave function renormalization factor $Z(s)$.
%
%By attaching $Z(s)$ to the potential sector,
%we get the renormalized G function as
To renormalize (\ref{G_molecule1}),
$Z(s)$ can be attached to the interaction sector
by $\bar{G}^{-1}\equiv {\rm diag}(\sqrt{Z}, 1)G^{-1}{\rm diag}(\sqrt{Z}, 1)$ as
\begin{eqnarray}
\hspace*{-1.5mm}\bar{G}^{-1}\!\!=\!\!\left(\begin{array}{cc}
                                         s-s_p & 0 \\
                                                 0  & s-m_{a_1}^2
                                          \end{array}\right)\!-\!
                                         \left(\begin{array}{cc}
                                                                                 0 & \sqrt{Z} v_{a_1\pi\rho} \\
                                                 \sqrt{Z}  v_{a_1\pi\rho} & 0
                                          \end{array}\right), \label{Gform3}
\end{eqnarray}
where the first term is the inverse of the  ``free''
propagator for the $\pi\rho$-molecule state and the $q\bar{q}$ state as the $a_1$ meson.
Now, by solving the 
%Klein-Gordon (KG) 
relativistic eigenvalue
equation for $\bar{G}$ as
\begin{eqnarray}
\det \bar{G}^{-1}=0, \label{KG1}
\end{eqnarray}
we have arrived at two-level model for the $a_1$ meson with
the $\pi\rho$-molecule and the $q\bar{q}$ components having proper mixing.

Now, to get the geometrical map on the complex-$N_c$ plane 
for the $a_1$ meson, 
%topologically see the relations between QCD with $N_c=3$
%and large-$N_c$ QCD with $N_c=\infty$,
we reduce Eq.~(\ref{KG1}) to the Schr\"{o}dinger equation
for Eq.~(\ref{H_1}), with a non-relativistic approximation.
We approximate the molecule propagator and 
the renormalization factor in Eq.~(\ref{G_molecule1})
as 
$(s-s_p)^{-1}\simeq \ltk 2\sqrt{s_p}(E-\sqrt{s_p})\rtk^{-1}$
and 
$\sqrt{Z}\simeq 84-21i$ estimated 
at $\sqrt{s}=\sqrt{s_p}\simeq 1012-221i$ in MeV unit.
We also approximate the $q\bar{q}$ propagator and the coupling constant 
as 
$(s-m_{a_1}^2)^{-1}\simeq \ltk 2m_{a_1}(E-m_{a_1})\rtk^{-1}$
and 
$v_{a_1\pi\rho}\simeq -6493$ at $\sqrt{s}=m_{a_1}=1189$ in MeV unit. 
Such energy fixing has been traditionally employed 
in nuclear-physics shell-model study,
where the absorptive effects into decay channels outside of the model space
are represented by the non-Hermite matrix elements~\cite{Romo}.
Then, 
Eq.~(\ref{KG1}) can be written as
\begin{eqnarray}
(E-\sqrt{s_p})(E-m_{a_1})-\frac{1}{(2\tilde{m})^2}(\sqrt{Z}v_{a_1\pi\rho})^2=0,\label{nonrela1}
\end{eqnarray}
with 
%a typical energy scale
$\tilde{m}\equiv \sqrt{\sqrt{s_p}m_{a_1}}$.
From the Schr\"{o}dinger equation (\ref{nonrela1}),
we can construct the two dimensional Hamilton matrix as
\begin{eqnarray}
H=
                  \left(\begin{array}{cc}
                   \sqrt{s_p} &\frac{1}{2\tilde{m}} \sqrt{Z} v_{a_1\pi\rho} \\
                    \frac{1}{2\tilde{m}} \sqrt{Z} v_{a_1\pi\rho}   & m_{a_1}
                             \end{array}\right). \label{Hnonrela}
\end{eqnarray}

%-----------------

Now we evaluate $N_c$-counting for the matrix elements in Eq.~(\ref{Hnonrela}).
According to large-$N_c$ QCD~\cite{tH, Witten},
$m_{a_1}$, $v_{a_1\pi\rho}$ and $G_{\pi\rho}$~\cite{comment1} have $N_c$-dependence as
\begin{eqnarray}
m_{a_1}\sim O(N_c^0), \hspace{2mm}v_{a_1\pi\rho}\sim O(1/\sqrt{N_c}), \hspace{2mm}G_{\pi\rho}\sim O(N_c^0).\label{count1}
\end{eqnarray}
For energy region far from the threshold; $s \gg (m_\rho+m_\pi)^2$,
the Weinberg-Tomozawa interaction (\ref{v_WT}) can be simplified as
$v_{\rm WT}\sim s \times O(1/N_c)$ as the mesonic four-point interaction~\cite{tH,Witten}.
Therefore, Eq.~(\ref{G_molecule1}) can be rewritten as
\begin{eqnarray}
%\frac{G_{\pi\rho}}{1-v_{\rm WT}G_{\pi\rho}}
G_{\rm WT}
\sim\frac{G_{\pi\rho}}{1-\{s\times O(1/N_c)\} G_{\pi\rho}}
\sim\frac{O(N_c)}{s-O(N_c)/G_{\pi\rho}}.\label{WTmod}
\end{eqnarray}
By comparing Eqs.(\ref{G_molecule1}) and (\ref{WTmod}),
we can also estimate the $N_c$ dependence of $\sqrt{s_p}$ and $\sqrt{Z}$  as
\begin{eqnarray}
\sqrt{s_p}\sim O(\sqrt{N_c}), \hspace{3mm}
\sqrt{Z}\sim O(\sqrt{N_c}), \label{count2}
\end{eqnarray}
where energy dependence of the loop function $G_{\pi\rho}$ is approximately 
ignored.
Such increasing behavior of $\sqrt{s_p}$ with $N_c$ as in Eq.~(\ref{count2})
can also be observed in the second reference of~\cite{PR}.
By using 
Eq.~(\ref{count2}),
we can also estimate the $N_c$-dependence of energy scale $\tilde{m}$ 
introduced in Eq.~(\ref{nonrela1}) as
\begin{eqnarray}
\tilde{m}\sim O(\sqrt[4]{N_c}). \label{count3}
\end{eqnarray}
%
%By using the $N_c$-countings (\ref{count1}), (\ref{count2}) and (\ref{count3})
By using Eqs.~(\ref{count1}), (\ref{count2}) and (\ref{count3})
for the matrix elements in Eq.~(\ref{Hnonrela}),
we eventually get the 
Complex 2D Matrix Model for $a_1$ meson with
$N_c$ dependence factored out by $\lambda$ as
\begin{eqnarray}
H(\lambda)=
                  \left(\begin{array}{cc}
                   \frac{1}{\lambda^2}\sqrt{s_p} &  \frac{\lambda}{2\tilde{m}}\sqrt{Z}v_{a_1\pi\rho} \\
                     \frac{\lambda}{2\tilde{m}} \sqrt{Z}v_{a_1\pi\rho}  & m_{a_1}
                             \end{array}\right), \label{H_C2Da1}\\
\lambda\equiv \sqrt[4]{3/N_c},\hspace{15mm}
\end{eqnarray}
where $\sqrt{s_p}$, $\sqrt{Z}$, $v_{a_1\pi\rho}$ and $\tilde{m}$
in (\ref{H_C2Da1}) are the constants estimated at $N_c=3$ 
as shown above Eq.~(\ref{nonrela1}).
$\frac{1}{\lambda^2}\sqrt{s_p}$ and $m_{a_1}$ in Eq.~(\ref{H_C2Da1})
are the energies of the $\pi\rho$-molecule state and the $q\bar{q}$ state, 
and they form the appropriate basis.
The (1,1) element with negative power of $\lambda$
reflects that a resonance state appears due to highly nonperturbative
hadron dynamics.
%
%\corr{In fact, the building block (\ref{H_C2Da1}) with the $\lambda$ dependence 
%is {\it universal} for any hadron with admixed natures of the mesonic molecule and the $q\bar{q}$
%components.}{}

%---------------

\begin{figure}[t]
  \begin{center}
  \hspace*{-2mm}
       \resizebox{100mm}{!}{\includegraphics{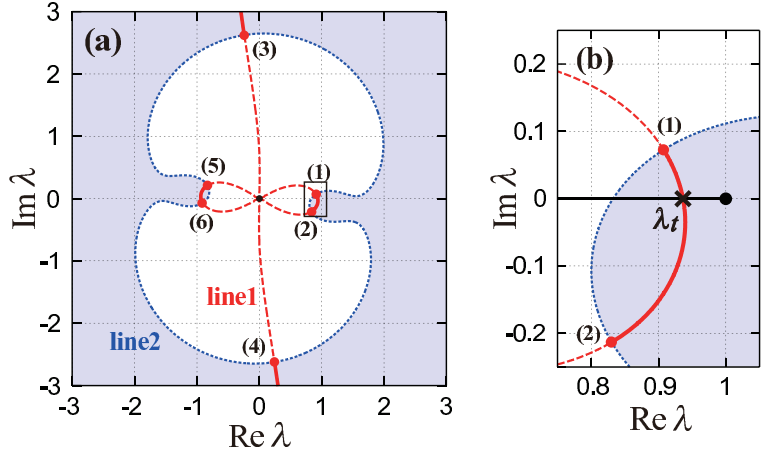}}\\
  \end{center}
\vspace{-2.2mm}
\caption{(Color) (a) Geometrical map
on the complex-$N_c$ plane with $\lambda=\sqrt[4]{3/N_c}$.
Constants in Eq.~(\ref{H_C2Da1}) are
$\sqrt{s_p}=1012-221i$,
$m_{a_1}=1189$,
$\sqrt{Z}=84-21i$ and
$v_{a_1\pi\rho}=-6493$
in MeV unit. 
Line 1 and shaded area with the boundary of line 2
correspond to the conditions (\ref{condF1}) and (\ref{condF2}),
respectively.
Six exceptional points $(n)$ $(n=1\sim 6)$
as the crossing points between line 1 and line 2,
and four transition lines as solid curves appear.
(b) Close-up figure around 
a blank square in (a). 
Transition line as a solid curve
crosses the real axis at $\lambda_t\sim 0.93$, i.e., $N_c\sim 4.0$,
which locates between $\lambda=0$ ($N_c=\infty$)
and $\lambda=1$ ($N_c=3$).
}
  \label{fig_2}
\end{figure}

Then, by applying the conditions (\ref{condF1}) and (\ref{condF2})
to the Hamiltonian (\ref{H_C2Da1}),
we can get the geometrical map on the complex-$N_c$ plane for the $a_1$ meson in Fig.~\ref{fig_2}.
%(See Appendix~\ref{A2} for a prescription of writing geometrical maps.)
(A prescription of writing geometrical maps is shown in Appendix~\ref{A2}.)
Six exceptional points and four transition lines,
two of which are half-lines,
appear on this map.
These numbers can be derived from
the power counting about $\lambda$
in Eq.~(\ref{EXcond1}).
The transition line shown by the solid curve 
can cross the real $\lambda$ axis between
$\lambda=0$ ($N_c=\infty$) and $\lambda=1$ ($N_c=3$).
The crossing point shows a critical color number
for transition as $\lambda_t=\sqrt[4]{3/N_c}\sim 0.93$, i.e., $N_c\sim 4.0$.
%
%\corr{In Fig.~\ref{PQR_fig}, we have discussed the 
%power-independence of the geometrical map around $\lambda=1$.
%Therefore, even if the estimation of $N_c$-countings in Eq. (\ref{H_C2Da1})
%were modified, the critical value $\lambda_t=(3/N_c)^{1/4}\sim 0.93$
%should not be affected so much.}{}
%
This result indicates that, with continuous change of $N_c$ from
$\infty$ to $3$,
the internal structures of two hadronic states can be exchanged
in terms of appropriate basis $q\bar{q}$ and $\pi\rho$-molecule at the 
critical color number $N_c\sim 4.0$.
Such a critical color number with character exchange for the $a_1$ meson
is also reported from the analysis of 
the pole residues in Ref.~\cite{NNOJH}.
In this way,
by looking into the existence of nature transition 
from the geometry on the complex-$N_c$ plane,
we can 
%\corr{successfully know the internal structure of hadrons
%with variation of $N_c$ from $\infty$ to $3$}{
discuss the typical $N_c$-dependence of the internal structure of hadrons
from $N_c=\infty$ to $3$.

\begin{figure}[t]
  \begin{center}
       \resizebox{155mm}{!}{\includegraphics{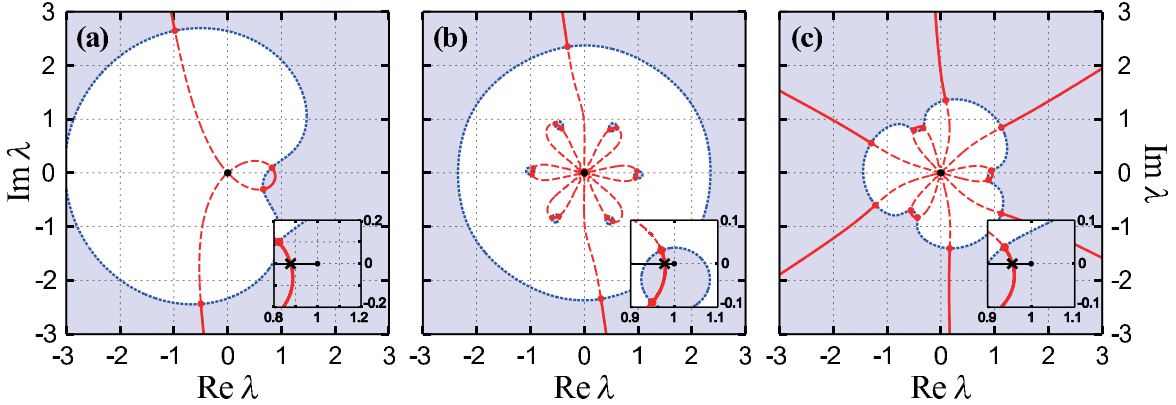}}\\
  \end{center}
\vspace{-2.2mm}
\caption{(Color) 
Geometrical maps on the complex-$N_c$ plane for 
(a) $H_{-101}$, 
(b) $H_{-601}$ and 
(c) $H_{-303}$  
with same constants of Fig.~\ref{fig_2}.
Notations for line 1, line 2, exceptional points and transition lines
are also same as those in Fig.~\ref{fig_2}. 
}
  \label{Nc_maps}
\end{figure}

Finally, we check 
the stability of the present results, especially that of the appearance of the critical
color number near $N_c=3$, 
%to some expected 
when corrections are made in the $N_c$-counting
of Eq.~(\ref{H_C2Da1}).
Below Eq.~(\ref{Hnonrela}), we have adopted two approximations:
i) $v_{\rm WT}\sim s\times O(N_c^{-1})$ for $s \gg (m_\rho+m_\pi)^2$, and
ii) neglecting the energy dependence of $G_{\pi\rho}$.
These two are introduced to 
perform a simple $N_c$-counting of the
matrix elements in Eq.~(\ref{H_C2Da1}) as a first study.
By including effects of their energy dependence,
the counting in Eq.~(\ref{H_C2Da1}) 
can be moderately changed.
First, Eq.~(\ref{H_C2Da1}) can be classified as $H_{-201}$
of the $PQR$ model~(\ref{PQRmodel}).
Now, there are three physical constraints;
I)~$P<0$: mass and decay width of molecule resonance should
increases at large-$N_c$,
II)~$Q=0$: mass of $q\bar{q}$ is independent of $N_c$, and
III)~$R>0$: QCD should become a free meson theory at large-$N_c$~\cite{tH, Witten}.
With these constraints,
we have checked various types of geometrical maps on the complex-$N_c$ plane
for physical sets of $P$, $Q$ and $R$ (For example, see Fig.~\ref{Nc_maps} for
$H_{-101}$, $H_{-601}$ and $H_{-303}$ as test cases).
We find that the appearance of the critical color number near
$N_c=3$ in this paper
is not so much affected by the corrections of
%relative correction of 
the $N_c$-counting in Eq.~(\ref{H_C2Da1}).
This is reasonably expected because $N_c=3$ corresponds to the fixed point on
the complex-$N_c$ plane as discussed in
the $PQR$ model (\ref{PQRmodel}).

%---------------------------------------------------------------------

%-----------------------------------------------------------------------

%-----------------------------------------------------------------------

%-----------------------------------------------------------------------
\section{Summary and outlook \label{Sum}}
%-------------------------------------------------------------------------------------------
We have formulated the Complex 2D Matrix Model
to get typical features about the parameter dependence of the internal structure of resonances.
We suggest that the geometry on the complex-parameter space
will give a simple criterion for the nature transition between resonance states
within the real-parameter
subspace. 
By applying the model to hadron physics,
we have discussed the $N_c$-dependence of the internal structure of hadrons 
from the geometry on the complex-$N_c$ plane.
We show that, 
with continuous change of $N_c$ from
$\infty$ to $3$,
the internal structures of hadrons can be exchanged
in terms of appropriate basis 
at the critical color number.
We hope that the new concept of geometry on the complex-$N_c$
plane and its possible topological classification
will shed light on the exotic physics in QCD for the future.

Our model can be employed to general multi-level problems of resonances
to analyze their internal structures with variation of a parameter in each system.
Wide applications of our model to resonance physics are expected
as a future prospect.

\section*{Acknowledgment}
The authors thank Hiroki Nakamura, Koichi Yazaki, and Tetsuo Hyodo for their fruitful communications.
The authors also thank Masuo Suzuki 
%and Naomichi Hatano
for his valuable suggestions
during the international workshop 
``Resonances and non-Hermitian systems in quantum mechanics (2012)''
in the Yukawa Institute for Theoretical Physics at Kyoto
University.
K.~N. thanks Naomichi Hatano for his meaningful discussions 
about probabilistic interpretation of resonances.
This work is supported by 
Grant-in-Aid for Scientific Research on Innovative Areas
``Elucidation of
New Hadrons with a Variety of Flavors''
(Nos. 22105509 (K. N.), 
22105510 (H. N.), 24105706 (D. J.) and E01:21105006 (A. H.))
from the Ministry of Education,
Culture, Sports, Science, and Technology(MEXT) of Japan.
K. N. is supported by the Special Postdoctoral Research Program
of RIKEN.

\appendix
%%%%%%%%%%%%%%%%%%%%%%%%%%%%%%%%%%%%%%
\section{Attaching number $N_{\rm at}$ \label{A1}}
%%%%%%%%%%%%%%%%%%%%%%%%%%%%%%%%%%%%%%
In this Appendix,
we show line 1 always crosses the point $\lambda=0$ for $R>0$
in the $PQR$ model. First,
we calculate $N_{\rm at}$ showing
how many times line 1 attaches to $\lambda=0$.
Eq.~(\ref{condF1}) can be written in the $PQR$ model (\ref{PQRmodel}) as
\begin{eqnarray}
{\rm Re}[A(\lambda)^*\overline{V}(\lambda)]=
\frac{1}{2}{\rm Re}[(\Delta\varepsilon+
\lambda^P v_{11}-\lambda^Q v_{22})^\ast \lambda^R \bar{v}]=0,\label{condF1_PQR}
\end{eqnarray}
with $\Delta\varepsilon\equiv\varepsilon_1^{(0)}-\varepsilon_2^{(0)}$
and $\bar{v}^2\equiv v_{12}v_{21}$. 
We define $N_A\equiv{\rm min}(0,P,Q)$ 
to find a largest contribution within $A(\lambda)^\ast$ 
at $\lambda \rightarrow 0$.
Then, at $\lambda \rightarrow 0$,
Eq.~(\ref{condF1_PQR}) becomes
\begin{eqnarray}
{\rm Re}[C(\lambda^{N_A})^\ast \lambda^R]=
ar^{R+N_A}{\rm Re}\ldk e^{i\{(R-N_A)\theta+\phi\}}\rdk=0,\label{condF1_N_A}
\end{eqnarray}
with $C\equiv ae^{i\phi}$ and $\lambda\equiv re^{i\theta}$.
Now there are two cases as follows.

1)$R-N_A=0$:
By definition,
$N_A\leq 0$ so that $R\leq 0$ and $r^{R+N_A}$
becomes unity or divergent at $\lambda\rightarrow 0$.
Therefore, Eq.~(\ref{condF1_N_A}) is not satisfied at $\lambda=0$,
i.e., $N_{\rm at}=0$.
Here we implicitly exclude the accidental case
$\phi=\frac{\pi}{2}+n\pi$ $(n\in {\bf Z})$ 
which makes the equation indefinite.

2)$R-N_A\neq 0$:
If the argument {\it strongly} approaches to certain angles at $\lambda\rightarrow 0$ as
\begin{eqnarray}
(R-N_A)\theta+\phi \rightarrow \frac{\pi}{2}+n\pi, 
\hspace{6mm} (n\in {\bf Z}) \label{angle1}
\end{eqnarray}
Eq.~(\ref{condF1_N_A}) is satisfied
instead of possible divergence of $r^{R+N_A}$ for $R+N_A<0$.
%
%Condition~
(\ref{angle1}) is equal to
\begin{eqnarray}
\theta \rightarrow \frac{\frac{\pi}{2}-\phi}{R-N_A}+n\frac{\pi}{R-N_A},
%\theta \rightarrow \frac{\frac{\pi}{2}-\phi}{R-N_A}+n\Delta\theta.
\hspace{6mm} (n\in {\bf Z}) \label{angle2}
\end{eqnarray}
Therefore
%so that
%Then,
$\Delta\theta\equiv\frac{\pi}{|R-N_A|}$ 
is an angle between nearest-neighboring two segments
of line 1 near $\lambda=0$.
$N_{\rm at}$ 
%can be derived as
is given by
\begin{eqnarray}
N_{\rm at}=\frac{2\pi}{\Delta\theta}=2|R-N_A|=2|R-{\rm min}(0,P,Q)|.\label{Nat_form1}
\end{eqnarray}
In table I, we summarize $N_{\rm at}$ for various sets of $(P,Q,R)$
discussed in Figs.~\ref{fig_1},~\ref{PQR_fig},~\ref{fig_2} and \ref{Nc_maps}.
$$\begin{array}{ccccccccc}
\multicolumn{8}{c}{\mbox{TABLE I: $N_{\rm at}$ for various sets of $(P,Q,R)$} }\\ \hline\hline
 &H_{111}&H_{221}&H_{233}&H_{151}&H_{-201}&H_{-101}&H_{-601}&H_{-303} \\
 \hline
N_{\rm at} &2&2&6&2&6&4&14&12\\\hline\hline
\end{array}$$
$N_{\rm at}$ in table I
is consistent with the resultant geometry of line 1 
around $\lambda=0$
in Figs.~\ref{fig_1},~\ref{PQR_fig},~\ref{fig_2} and \ref{Nc_maps}.

Now, if $R>0$ consistently with $V_{ij}(0)=0$ in Eq.~(\ref{H_1}),
$R-N_A\neq 0$ because $N_A\leq 0$.
Therefore $N_{\rm at}>0$ and line 1 always crosses $\lambda=0$.

%%%%%%%%%%%%%%%%%%%%%%%%%%%%%%%%%%%%%%
\section{Prescription of geometrical maps \label{A2}}
%%%%%%%%%%%%%%%%%%%%%%%%%%%%%%%%%%%%%%

%-----------------------------------------------------------------------

\begin{figure}[t]
  \begin{center}
       \resizebox{120mm}{!}{\includegraphics{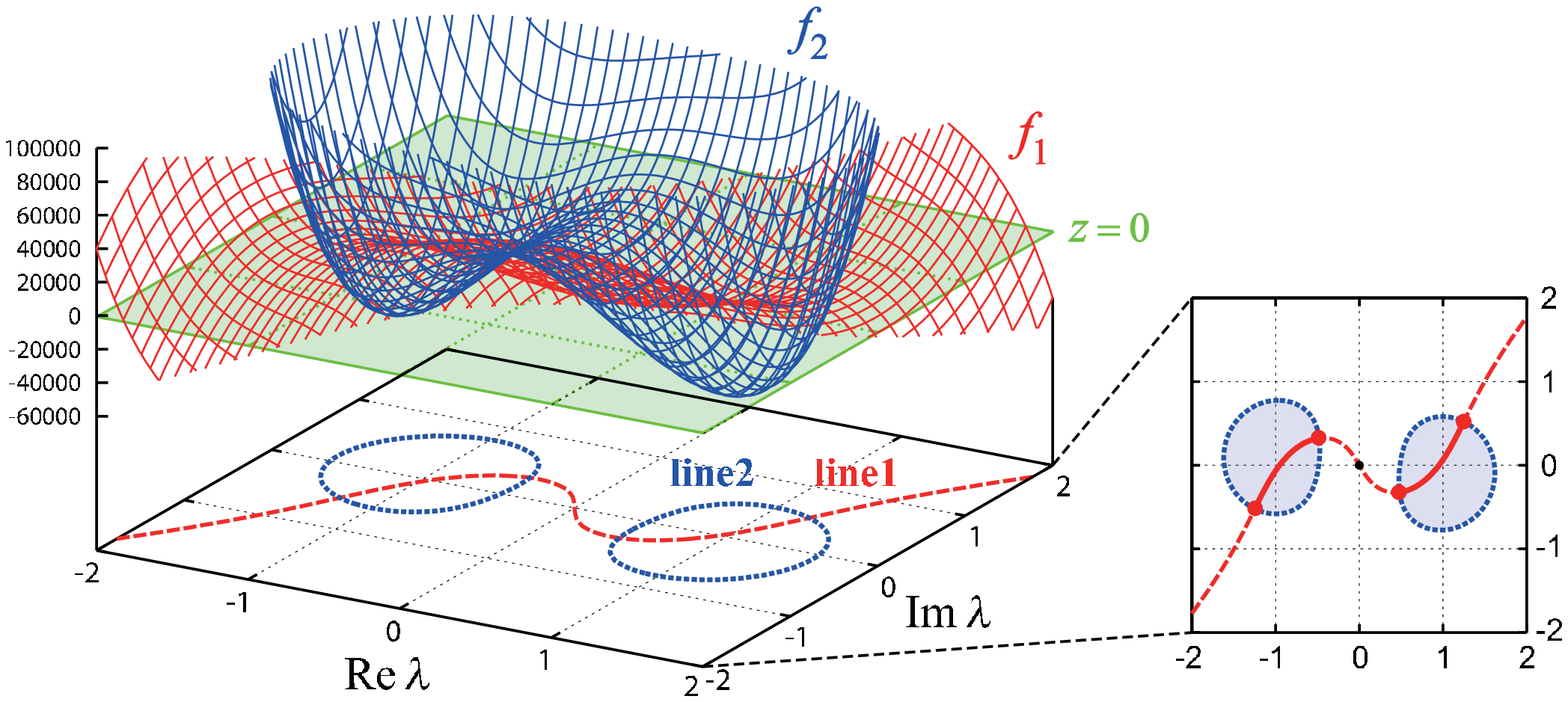}}\\
  \end{center}
%\vspace{-4mm}
\caption{(Color) Three dimensional plot of $z=f_i(x,y)$ $(i=1,2)$ with
contours at $z=0$ (Left) and corresponding geometrical map (Right).
Constants and powers of the matrix elements 
are same as those of Fig.~\ref{PQR_fig}(a).
}
\label{3dim_plot}
\end{figure}
%-----------------------------------------------------------------------

In this Appendix, we show a prescription of writing geometrical maps.
Eq.~(\ref{condF1}) and a boundary of Eq.~(\ref{condF2})
can be rewritten as algebraic equations:
\begin{eqnarray}
f_i({\rm Re}\lambda, {\rm Im}\lambda)=0, \label{algebra1}
\end{eqnarray}
where $f_i$ is a real function of $\lambda \in {\bf C}$
and $i=1$ ($i=2$) corresponds to Eq.~(\ref{condF1}) 
(boundary of Eq.~(\ref{condF2})).
By drawing a three-dimensional plot of $z=f_i(x,y)$,
{\it contours} at $z=0$ correspond to line 1 (for $i=1$)
and line 2 (for $i=2$) (See, e.g., Fig.~\ref{3dim_plot}).
In this way, by using a contour method,
one can get geometrical maps
without numerically solving the high-powered algebraic equations (\ref{condF1}) and (\ref{condF2}).


\begin{thebibliography}{99}
\bibitem{comment1}
If $N_c$-dependent cut-off~\cite{PR} is considered,
$G_{\pi\rho}$ would have 
complicated $N_c$-dependence  (even in a
non-analytic way).
In our present study, to simply perform the analytic continuation
into the complex-$N_c$ plane,
we alternatively use
$N_c$-independent cut-off 
%as another candidate~\cite{PR}
%
corresponding to  the mass scale of $q\bar{q}$ meson,
which is also  suggested as another candidate in Ref.~\cite{PR}.
\end{thebibliography}
\end{document}